# A Methodology for Empirical Quality Assessment of Object-Oriented Design

Devpriya Soni[1]
*Department of Computer Applications*

Dr. Namita Shrivastava[2]
*Asst. Prof. Deptt. of Mathematics*

Dr. M. Kumar[3]
*Retd. Prof. of Computer Applications*

*Maulana Azad National Institute of Technology (A Deemed University)*
*Bhopal 462007, India*

*Abstract:* **The direct measurement of quality is difficult because there is no way we can measure quality factors. For measuring these factors, we have to express them in terms of metrics or models. Researchers have developed quality models that attempt to measure quality in terms of attributes, characteristics and metrics. In this work we have proposed the methodology of controlled experimentation coupled with power of Logical Scoring of Preferences to evaluate global quality of four object-oriented designs.**

## I. INTRODUCTION

Software quality must be addressed during the whole process of software development. However, design is of particular importance in developing quality software for two reasons: (i) design is the first stage in software system creation in which quality requirement can begin to be addressed. Error made at this stage can be costly, even impossible to be rectified. (ii) design decision has significant effect on quality on the final product.

Measuring quality in the early stage of software development is the key to develop high-quality software. Analyzing object-oriented software in order to evaluate its quality is becoming increasingly important as the paradigm continues to increase in popularity. A large number of software product metrics have been proposed in software engineering. While many of these metrics are based on good ideas about what is important to measure in software to capture its complexity, it is still necessary to systematically validate them. Recent software engineering literature has shown a concern for the quality of methods to validate software product metrics (e.g., see [1][2][3]). This concern is due to fact that: (i) common practices for the validation of software engineering metrics are not acceptable on scientific grounds, and (ii) valid measures are essential for effective software project management and sound empirical research. For example, Kitchenham et.al. [2] write: "Unless the software measurement community can agree on a valid, consistent, and comprehensive theory of measurement validation, we have no scientific basis for the discipline of software measurement, a situation potentially disastrous for both practice and research."

According to Fenton [4], there are two types of validation that are recognized: internal and external. Internal and external validations are also commonly referred to as theoretical and empirical validation respectively [2]. Both types of validation are necessary. Theoretical validation requires that the software engineering community reach a consensus on what are the properties for common software maintainability metrics for object-oriented design. Software organizations can use validated product metrics in at least three ways: to identify high risk software components early, to construct design and programming guidelines, and to make system level predictions. Empirical validation can be performed through surveys, experiments and case-study.

Recently, Kumar and Soni [5] have proposed a hierarchical model to evaluate quality of object-oriented software. The proposed model of [5] has been validated both theoretically as well as empirically in a recent paper by Soni, Shrivastava and Kumar [6]. Further the model has been used for evaluation of maintainability assessment of object-oriented design quality, especially in design phase, by Soni and Kumar [7]. In this research, the authors have attempted to empirically validate the object-oriented design model of [5] using the methodology of controlled experiment. A global quality assessment of several designs have been made using the method of Logical scoring of Preferences (LSP). The Section II deals with experimental environment and data collection and the Section III deals with the method of Logical Scoring of Preferences (LSP) used to evaluate the overall quality of software design. Section IV gives the steps for design quality evaluation and Section V analyzes and compare the quality of selected designs.

*(IJCSIS) International Journal of Computer Science and Information Security,*
*Vol. 7, No.2, 2010*## II. EXPERIMENTAL ENVIRONMENT AND DATA COLLECTION

For the purpose of empirically evaluating object-oriented design for its quality using the hierarchical quality model proposed by Kumar and Soni [5], we needed a few designs created independently for the same problem/project. We used 12 students of fifth semester, Master of Computer Applications of Maulana Azad National Institute of Technology, Bhopal. They had studied courses on Data Base Management System, Object-Oriented Analysis and Design and C++ programming language course including laboratory on these topics. We formed three groups of 4 students each. These groups were provided a written problem statement (user requirements) for designing a small sized library management system for MANIT library. For any difficulty they were free to consult library staff. The three groups independently created one design each for the library management system. They were asked to follow Object-Oriented Analysis and Design methodology [8] for designing and were given two months to complete the work and produce design using methodology of discussion and walk-through within its group. The three designs produced are given in Fig 13, 14 and 15 (see Appendix A). To make this work more reliable and trustworthy, we also evaluated an object-oriented design of Human Resource Department [13]. This design was used to raise HR database, which is being successfully used by Bharat Heavy Electrical Limited (BHEL), Bhopal. This design is produced in Fig 16 (see Appendix A).

## III. LOGICAL SCORING OF PREFERENCES METHOD

The Logical Scoring of Preferences (LSP) method was proposed in 1996 by Dujmovic [9][11][12] who used it to evaluate and select complex hardware and software systems. It is grounded on Continuous Preference Logic. In LSP, the features are decomposed into aggregation blocks. This decomposition continues within each block until all the lowest level features are directly measurable. A tree of decomposed features and sub-factors at one level will have a number of aggregation blocks, each resulting in a higher-level factors going up the tree right through to the highest-level features. For each feature, an elementary criterion is defined. For this, the elementary preference $E_i$ needs to be determined by calculating a percentage from the feature score $X_i$. This relationship is represented in the following equation:

$$E_i = G_i(X_i) \qquad (1)$$

where E is the elementary preference, G is the function for calculating E, X is the score of a feature and i is the number of a particular feature. The elementary preferences for each measurable feature in one aggregation block are used to calculate the preference score of the higher feature. This in turn is used with the preferences scores of an even higher feature, continuing right up until a global preference is reached. The global preference is defined as:

$$E = L(E_1 ..., E_n) \qquad (2)$$

where E is the global preference, L is the function for evaluating E, En is the elementary preference of feature n, n is the number of features in the aggregation block. The function L yields an output preference $e_0$, for the global preference E, or any subfeature $E_i$. It is calculated as:

$$e_0 = (W_1 E_1^r + ... + W_k E_k^r)^{1/r}, W_1 + ... + W_k = 1 \qquad (3)$$

where $e_0$ is the output preference, W is the weight of the particular feature, E is the elementary preference of a feature, k is the number of features in the aggregation block and r is a conjunctive/disjunctive coefficient of the aggregation block. For each $E_i$ a weight $W_i$ is defined for the corresponding feature. The weight is a fraction of 1 and signifies the importance of a particular feature within the aggregation block. The r coefficient represents the degree of simultaneity for a group of features within an aggregation block. This is described in terms of conjunction and disjunction. The modification of above model, called Logic Scoring of Preferences, is a generalization of the additive-scoring model and can be expressed as follows

$$P/GP(r) = (W_1 EP_1^r + W_2 EP_2^r + ... + W_m EP_m^r)^{1/r} \qquad (4)$$

where $W_i$ weights and $EP_i$ are elementary preferences. The power r is a parameter selected to achieve the desired logical relationship and polarization intensity of the aggregation function. Value of 'r' used in Logic Scoring of Preferences method is given in Table I.

TABLE I.   VALUE OF R IN LOGIC SCORING OF PREFERENCE METHOD

| Operation | Symbol | d | r2 | r3 | r4 | r5 |
|---|---|---|---|---|---|---|
| ARITHMETIC MEAN | A | 0.5000 | 1.000 | 1.000 | 1.000 | 1.000 |
| WEAK QC (-) | C-- | 0.4375 | 0.619 | 0.573 | 0.546 | 0.526 |
| WEAK QC (+) | C-+ | 0.3125 | -0.148 | -0.208 | -0.235 | -0.251 |

The strength of LSP resides in the power to model different logical relationships:

- Simultaneity, when is perceived that two or more input preferences must be present simultaneously

- Replaceability, when is perceived that two or more attributes can be replaced (there exist alternatives, i.e., a low quality of an input preference can always be compensated by a high quality of some other input).

- Neutrality, when is perceived that two or more input preferences can be grouped independently (neither conjunctive nor disjunctive relationship)

- Symmetric relationships, when is perceived that two or more input preferences affect evaluation in the same logical way (tough may be with different weights).

- Asymmetric relationships, when mandatory attributes are combined with desirable or optional ones; and when sufficient attributes are combined with desirable or optional ones.

## IV. STEPS FOR DESIGN QUALITY EVALUATION

Steps required for the evaluation of design quality are:

1. **Consider a hierarchical model for quality characteristics and attributes (i.e. $A_1 \ldots A_n$):** here, we define and specify the quality characteristics and attributes, grouping them into a model. For each quantifiable attribute Ai, we can associate a variable $X_i$, which can take a real value: the measured value.

This function is a mapping of the measured value in the empirical domain [10] into the new numerical domain. Then the final outcome is mapped in a preference called the elementary quality preference, $EQ_i$. We can assume the elementary quality preference $EQ_i$ as the percentage of

```
1  Functionality
 1.1  Design Size
  1.1.1  Number of Classes (NOC)
 1.2  Hierarchies
  1.2.1  Number of Hierarchies (NOH)
 1.3  Cohesion
  1.3.1  Cohesion Among Methods of Class (CAM)
 1.4  Polymorphism
  1.4.1  Number of Polymorphic Methods (NOP)
 1.5  Messaging
  1.5.1  Class Interface Size (CIS)

2  Effectiveness
 2.1  Abstraction
  2.1.1  Number of Ancestors (NOA)
  2.1.2  Number of Hierarchies (NOH)
  2.1.3  Maximum number of Depth of Inheritance
         (MDIT)
 2.2  Encapsulation
  2.2.1  Data Access Ratio (DAR)
 2.3  Composition
  2.3.1  Number of aggregation relationships
         (NAR)
  2.3.2  Number of aggregation hierarchies (NAH)
 2.4  Inheritance
  2.4.1  Functional Abstraction (FA)
 2.5  Polymorphism
  2.5.1  Number of Polymorphic Methods (NOP)

3  Understandability
 3.1  Encapsulation
  3.1.1  Data Access Ratio (DAR)
 3.2  Cohesion
  3.2.1  Cohesion Among Methods of Class (CAM)
 3.3  Inheritance
  3.3.1  Functional Abstraction (FA)
 3.4  Polymorphism
  3.4.1  Number of Polymorphic Methods (NOP)

4  Reusability
 4.1  Design Size
  4.1.1  Number of Classes (NOC)
 4.2  Coupling
  4.2.1  Direct Class Coupling (DCC)
 4.3  Cohesion
  4.3.1  Cohesion Among Methods of Class (CAM)
 4.4  Messaging
  4.4.1  Class Interface Size (CIS)

5  Maintainability
 5.1  Design Size
  5.1.1  Number of Classes (NOC)
 5.2  Hierarchies
  5.2.1  Number of Hierarchies (NOH)
 5.3  Abstraction
  5.3.1  Number of Ancestors (NOA)
 5.4  Encapsulation
  5.4.1  Data Access Ratio (DAR)
 5.5  Coupling
  5.5.1  Direct Class Coupling (DCC)
  5.5.2  Number of Methods (NOM)
 5.6  Composition
  5.6.1  Number of aggregation relationships
         (NAR)
  5.6.2  Number of aggregation hierarchies (NAH)
 5.7  Polymorphism
  5.7.1  Number of Polymorphic Methods (NOP)
 5.8  Documentation
  5.8.1  Extent of Documentation (EOD)
```

Figure 1 Proposed hierarchical design quality model

2. **Defining criterion function for each attribute, and applying attribute measurement:** In this process, we define the basis for elementary evaluation criteria and perform the measurement sub-process. Elementary evaluation criteria specifies how to measure quantifiable attributes. The result is an elementary preference, which can be interpreted as the degree or percentage of satisfied requirement. For each variable $X_i$, $i = 1, \ldots, n$ it is necessary to establish an acceptable range of values and define a function, called the elementary criterion.

requirement satisfied by the value of $X_i$. In this sense, $EQ_i = 0\%$ denotes a totally unsatisfactory situation, while $EQ_i = 100\%$ represents a fully satisfactory situation, Dujmovic (1996). Ultimately, for each quantifiable attribute, the measurement activity should be carried out.

3. **Evaluating elementary preferences:** In this task, we prepare and enact the evaluation process to obtain an indicator of partial preference for design. For n attributes, the mapping produces n elementary quality preferences.

4. **Analyzing and assessing partial and global quality preferences:** In this final step, we analyze and assess the elementary, partial and total quantitative results regarding the established goals.

### A. Establishing Elementary Criteria

For each attribute $A_i$ we associate a variable $X_i$ which can take a real value by means of the elementary criterion function. The final result represents a mapping of the function value into the elementary quality preference, $EQ_i$. The value of $EQ_i$ is a real value that 'fortunately' belongs to the unit interval. Further, the preference can be categorized in three rating levels namely: satisfactory (from 60 to 100%), marginal (from 40 to 60%), and unsatisfactory (from 0 to 40%). For instance, a marginal score for an attribute could indicate that a correction action to improve the attribute quality should be taken into account by the manager or developer. Figure 2, shows sample elementary criteria for attributes.

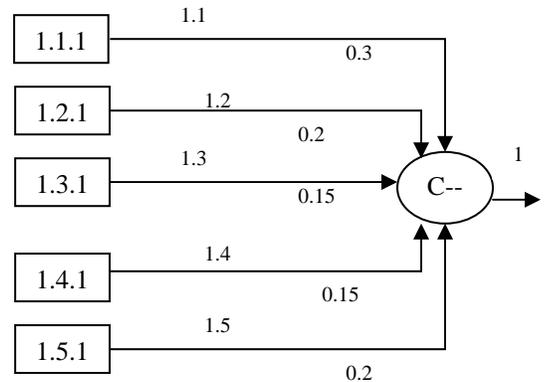

Figure 2 Sample elementary criteria defined as preference scales for the hierarchical model.

The preference scale for the *Number of Classes (NOC)* metric is a multi-level discrete absolute criterion defined as a subset, where 0 implies no classes available; 8 or more implies satisfactory (100%) number of classes present. The preference scale for the *Number of Hierarchies (NOH)* metric is a multi-level discrete absolute criterion defined as a subset, where 0 implies no hierarchy available; 5 or more implies satisfactory (100%) number of hierarchies present. The preference scale for the *Maximum Depth of Inheritance (MDIT)* metric is a multi-level discrete absolute criterion defined as a subset, where 0 implies depth is 1 level; 6 or more implies depth is satisfactory (100%).

The preference scale for the *Data Access Ratio (DAR)* metric is a multi-level discrete absolute criterion defined as a subset, where 0 implies ratio is less then 5%; 80% or more implies satisfactory (100%) ratio. The preference scale for the *Extent of Documentation (EOD)* metric is a multi-level discrete absolute criterion defined as a subset, where 0 implies that documentation present is 5% or less; 100% implies satisfactory (100%) documentation available. Similar criteria were followed for other metrics as well.

### B. Logic Aggregation of Elementary Preferences

Evaluation process is to obtain a quality indicator for each competitive system then applying a stepwise aggregation mechanism, the elementary quality preferences can be accordingly structured to allow the computing of partial preferences. Figure 3 to 7 depicts the aggregation structure for *functionality, effectiveness, understandability, reusability and maintainability*.

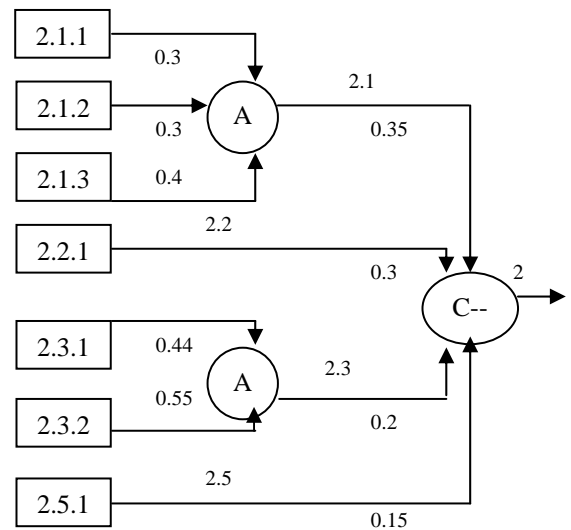

Figure 3 Structure of Partial Logic Aggregation for Functionality Factor

Figure 4 Structure of Partial Logic Aggregation for Effectiveness Factor

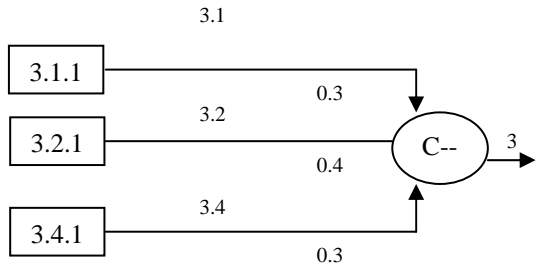

Figure 5 Structure of Partial Logic Aggregation for Understandability Factor

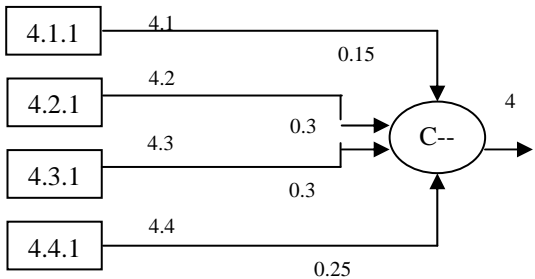

Figure 6 Structure of Partial Logic Aggregation for Reusability Factor

The global preferences can be obtained through repeating the aggregation process at the end. The global quality preference represents the global degree of satisfaction of all involved requirements. To evaluate the global quality it is necessary to assign elementary preference to each metric of the hierarchical model in Figure 1. Figure 8 shows the high-level characteristics aggregation to yield the global preference. The stepwise aggregation process follows the hierarchical structure of the hierarchical model from bottom to top. The major CLP operators are the arithmetic means (A) that models the neutrality relationship; the pure conjunction (C), and quasi-conjunction operators that model the simultaneity one; and the pure disjunction(D), and quasi-disjunction operators that model the replaceability one. With regard to levels of simultaneity, we may utilize the week (C-), medium (CA), and strong (C+) quasi-conjunction functions. In this sense, operators of quasi-conjunction are *flexible and* logic connectives. Also, we can tune these operators to intermediate values. For instance, C-- is positioned between A and C- operators; and C-+ is between CA and C operators, and so on. The above operators (except A) mean that, given a low quality of an input preference can never be well compensated by a high quality of some other input to output a high quality preference. For example in the Figure 3 at the end of the aggregation process we have the sub-characteristic coded 1.1 (called *Design Size* in the hierarchical Model, with a relative importance or weight of 0.3), *and* 1.2 sub- characteristic (*Hierarchies*, 0.2 weighted), *and* 1.3 sub-characteristic (*Cohesion*, 0.15 weighted), *and* 1.4 sub-characteristic (*Polymorphism*, 0.15 weighted), and 1.5 sub-characteristic (*Messaging*, 0.3 weighted).

All these sub-characteristic preferences are input to the C-- logical function, which produce the partial global preference coded as 1, (called *Functionality*).

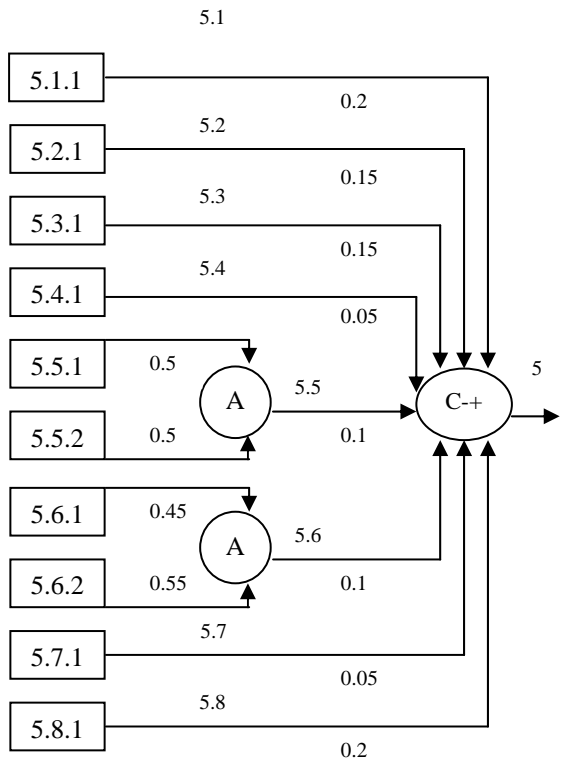

Figure 7 Structure of Partial Logic Aggregation for Maintainability Factor

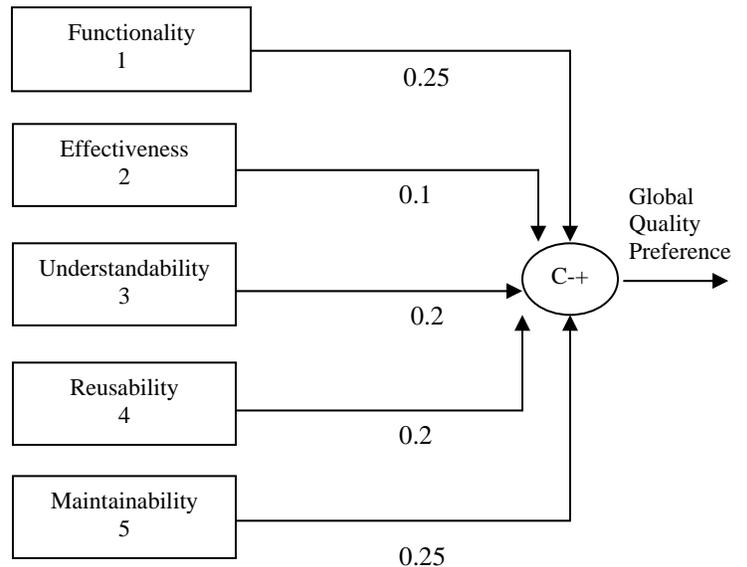

Figure 8 Global Aggregation of Preferences of Quality

## V. ANALYZING AND COMPARING THE QUALITY OF THE SELECTED DESIGNS

We have measured metrics values of all the four designs (shown in Appendix A in Fig. 13 to 16) and have worked out elementary preferences as discussed in the previous section. The results of partial quality preferences for *functionality*, *understandability*, *reusability*, *effectiveness* and *maintainability* of designs are shown in Table II to VI. A comparison of partial and global preferences of factors is given in Table VII for all the four designs. A bar chart representing the global quality of four designs is given in Fig 12.

TABLE II. PARTIAL QUALITY PREFERENCE FOR FUNCTIONALITY OF DESIGN

| Characteristics and Sub-characteristics | LMS -1 | LMS -2 | LMS -3 | HRIS |
|---|---|---|---|---|
| 1. Functionality | | | | |
| 1.1 Design Size | | | | |
| 1.1.1 Number of Classes (NOC) | 1 EQ=100% | 1 | 1 | 1 |
| 1.2 Hierarchies | | | | |
| 1.2.1 Number of Hierarchies (NOH) | .4 | .4 | .4 | .7 |
| 1.3 Cohesion | | | | |
| 1.3.1 Cohesion Among Methods of Class (CAM) | .8 | .7 | .6 | .8 |
| 1.4 Polymorphism | | | | |
| 1.4.1 Number of Polymorphic Methods (NOP) | 1 | 1 | 1 | .8 |
| 1.5 Messaging | | | | |
| 1.5.1 Class Interface Size (CIS) | .7 | .6 | .5 | .8 |
| **Partial Quality Preference** | **77.19** | **73.54** | **69.69** | **86.58** |

TABLE III. PARTIAL QUALITY PREFERENCE FOR UNDERSTANDABILITY OF DESIGN

| Characteristics and Sub-characteristics | LMS -1 | LMS -2 | LMS -3 | HRIS |
|---|---|---|---|---|
| 3. Understandability | | | | |
| 3.1 Encapsulation | | | | |
| 3.1.1 Data Access Ratio (DAR) | 1 | .8 | .6 | .8 |
| 3.2 Cohesion | | | | |
| 3.2.1 Cohesion Among Methods of Class (CAM) | .8 | .7 | .6 | .8 |
| 3.4 Polymorphism | | | | |
| 3.4.1 Number of Polymorphic Methods (NOP) | 1 | 1 | 1 | 1 |
| **Partial Quality Preference** | **91.77** | **81.60** | **71.08** | **85.79** |

TABLE IV. PARTIAL QUALITY PREFERENCE FOR EFFECTIVENESS OF DESIGN

| Characteristics and Sub-characteristics | LMS -1 | LMS -2 | LMS -3 | HRIS |
|---|---|---|---|---|
| 2. Effectiveness | | | | |
| 2.1 Abstraction | | | | |
| 2.1.1 Number of Ancestors (NOA) | .5 | .4 | .3 | .8 |
| 2.1.2 Number of Hierarchies (NOH) | .4 | .4 | .4 | .7 |
| 2.1.3 Maximum number of Depth of Inheritance (MDIT) | .5 | .4 | .2 | .6 |
| 2.2 Encapsulation | | | | |
| 2.2.1 Data Access Ratio (DAR) | 1 | .8 | .6 | .8 |
| 2.3 Composition | | | | |
| 2.3.1 Number of aggregation relationships (NAR) | .4 | .3 | .4 | .5 |
| 2.3.1 Number of aggregation hierarchies (NAH) | .8 | .7 | .6 | .7 |
| 2.5 Polymorphism | | | | |
| 2.5.1 Number of Polymorphic Methods (NOP) | 1 | 1 | 1 | 1 |
| **Partial Quality Preference** | **72.00** | **61.62** | **51.15** | **76.71** |

TABLE V. PARTIAL QUALITY PREFERENCE FOR REUSABILITY OF DESIGN

| Characteristics and Sub-characteristics | LMS -1 | LMS -2 | LMS -3 | HRIS |
|---|---|---|---|---|
| 4. Reusability | | | | |
| 4.1 Design Size | | | | |
| 4.1.1 Number of Classes (NOC) | 1 | 1 | 1 | 1 |
| 4.2 Coupling | | | | |
| 4.2.1 Direct Class Coupling (DCC) | 1 | 1 | 1 | 1 |
| 4.3 Cohesion | | | | |
| 4.3.1 Cohesion Among Methods of Class (CAM) | .8 | .7 | .6 | .8 |
| 4.4 Messaging | | | | |
| 4.4.1 Class Interface Size (CIS) | .7 | .6 | .5 | .8 |
| **Partial Quality Preference** | **86.06** | **80.12** | **73.97** | **88.75** |

TABLE VI. PARTIAL QUALITY PREFERENCE FOR MAINTAINABILITY OF DESIGN

| Characteristics and Sub-characteristics | LMS -1 | LMS -2 | LMS -3 | HRIS |
|---|---|---|---|---|
| **5. Maintainability** | | | | |
| 5.1 Design Size | | | | |
| 5.1.1 Number of Classes (NOC) | 1 | 1 | 1 | 1 |
| 5.2 Hierarchies | | | | |
| 5.2.1 Number of Hierarchies (NOH) | .4 | .4 | .4 | .7 |
| 5.3 Abstraction | | | | |
| 5.3.1 Number of Ancestors (NOA) | .5 | .4 | .3 | .8 |
| 5.4 Encapsulation | | | | |
| 5.4.1 Data Access Ratio (DAR) | 1 | .8 | .6 | .8 |
| 5.5 Coupling | | | | |
| 5.5.1 Direct Class Coupling (DCC) | 1 | 1 | 1 | 1 |
| 5.5.2 Number of Methods (NOM) | 1 | 1 | 1 | 1 |
| 5.6 Composition | | | | |
| 5.6.1 Number of aggregation relationships (NAR) | .4 | .3 | .4 | .5 |
| 5.6.2 Number of aggregation hierarchies (NAH) | .8 | .7 | .6 | .7 |
| 5.7 Polymorphism | | | | |
| 5.7.1 Number of Polymorphic Methods (NOP) | 1 | 1 | 1 | 1 |
| 5.8 Documentation | | | | |
| 5.8.1 Extent of Documentation (EOD) | .7 | .8 | .7 | .7 |
| **Partial Quality Preference** | **68.54** | **65.82** | **59.98** | **79.98** |

TABLE VII. QUALITY FACTORS AND GLOBAL QUALITY FACTORS OF VARIOUS DESIGNS

| Quality Factors → Design ↓ | Functionality | Effectiveness | Understandability | Reusability | Maintainability | Global Quality Preferences |
|---|---|---|---|---|---|---|
| LMS -1 | 77.19 | 72 | 91.77 | 86.06 | 68.54 | 78.61 |
| LMS -2 | 73.54 | 61.62 | 81.6 | 80.12 | 65.82 | 72.9 |
| LMS -3 | 69.69 | 51.15 | 71.08 | 73.97 | 59.98 | 66.01 |
| HRIS | 86.58 | 76.71 | 85.79 | 88.75 | 79.98 | 84.07 |

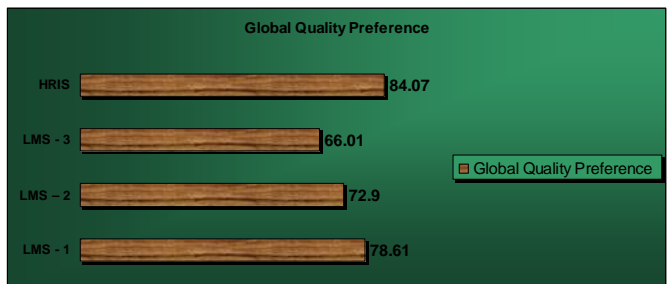

Figure 12 Global Quality of Designs

## VI. CONCLUSION

We have used the Logical Scoring of Preferences method to evaluate global quality of four designs, three created by fifth semester Master of Computer Applications students and the fourth one created by professionals. As expected the global quality index of design created by professionals has the highest quality index of 84.07 followed by design LMS-1, which has the value 78.61. We believe that the methodology used is quite simple and will provide reasonable estimates for factors like *functionality, effectiveness, reusability, understandability,* and *maintainability* and also the overall quality of software design. It is worth mentioning that a reasonable estimate of maintainability of software design is going to be very useful for software professionals.

Appendix A

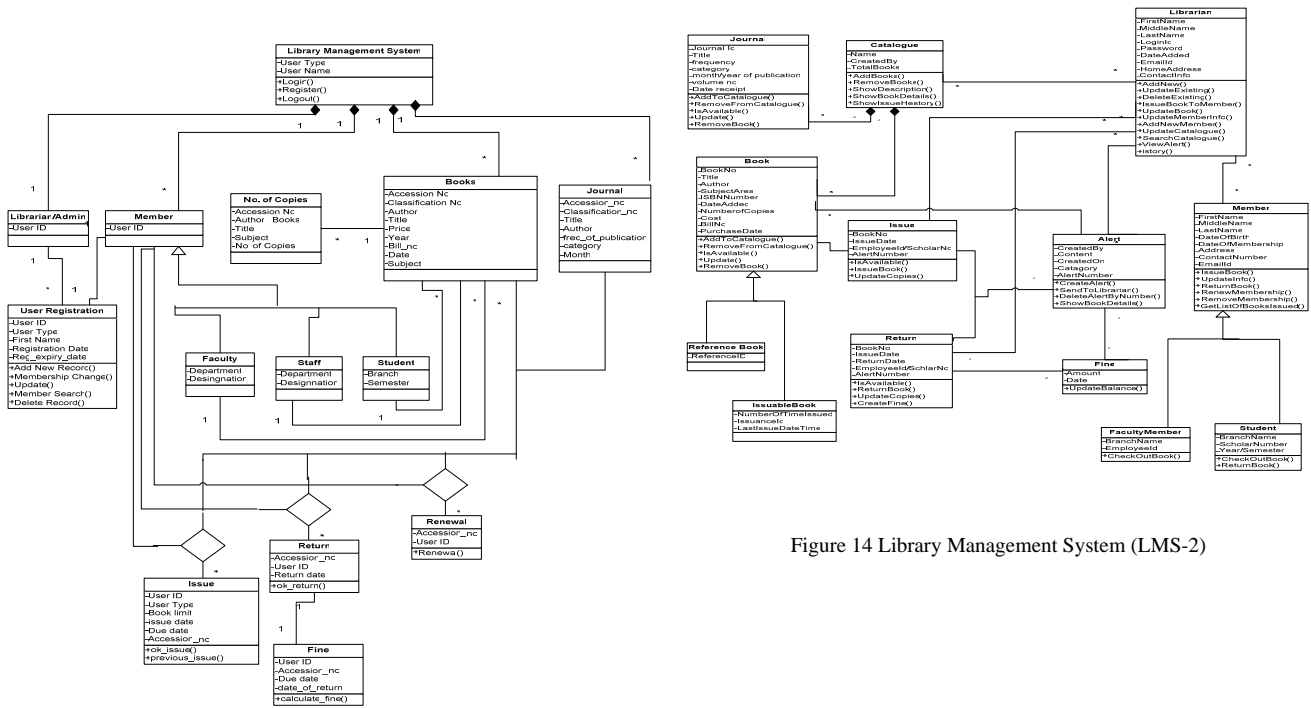

Figure 14 Library Management System (LMS-2)

Figure 13 Library Management System (LMS-1)

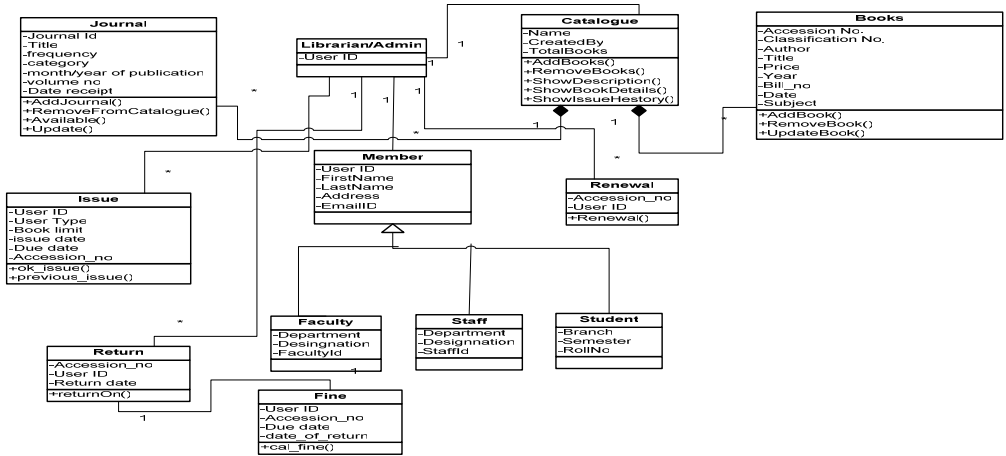

Figure 15 Library Management System (LMS-3)

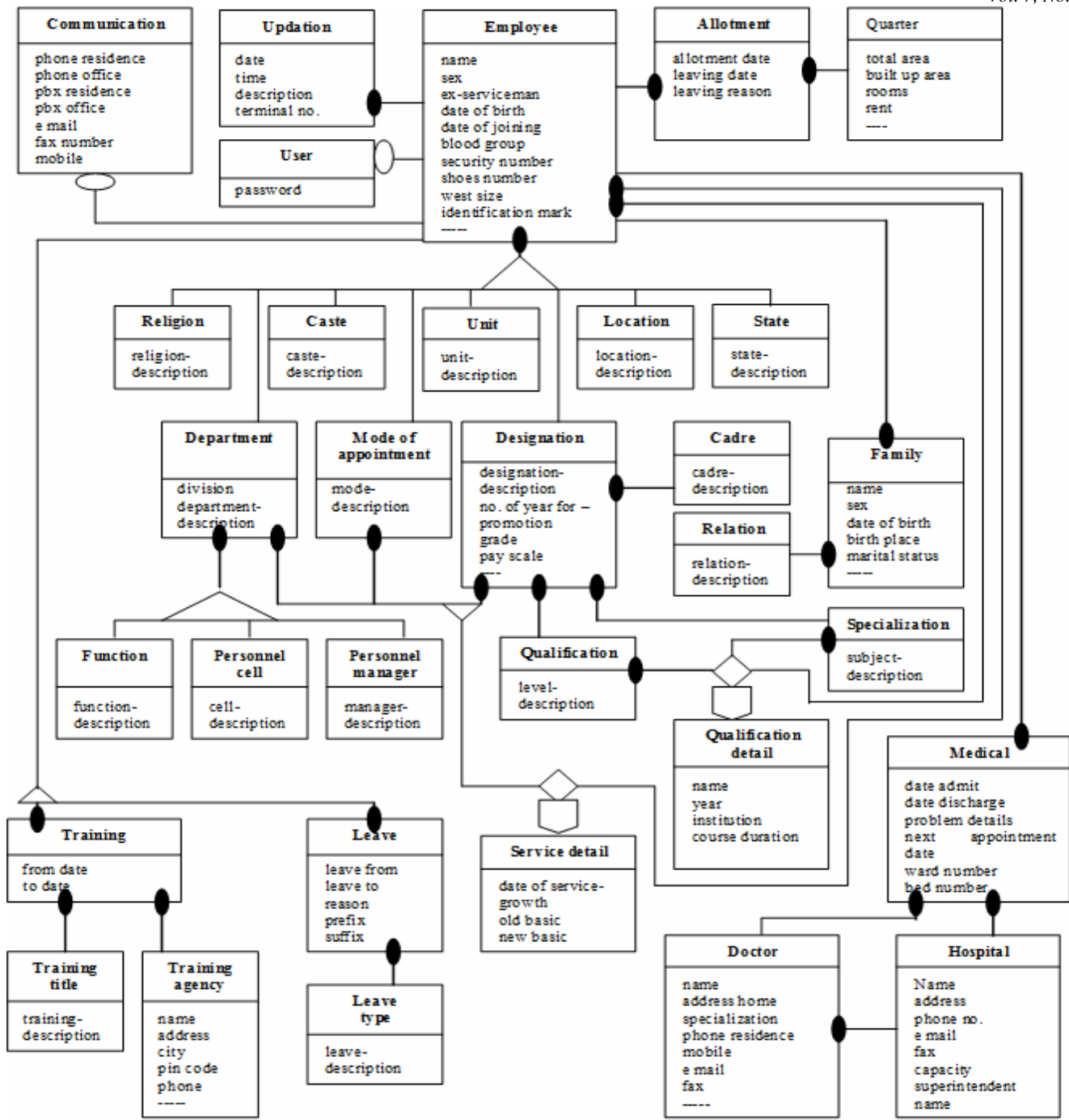

Figure 16 Human Resource Information System (HRIS)

AUTHORS PROFILE

**Devpriya Soni** has seven years of teaching experience to post graduate classes and four years of research experience at MANIT. She is pursuing her Ph.D. at Department of Computer Applications, MANIT, Bhopal. Her research interest is object-oriented metrics and object-oriented databases. EmailId: devpriyasoni@gmail.com

**Dr. Namita Shrivastava** has done M.Sc., Ph.D. She has 19 years of teaching and 18 years of research experience. Her area of interest is crack problem, data mining, parallel mining and object-oriented metrics. EmailId: sri.namita@gmail.com

**Dr. Mahendra Kumar** is presently Prof. & Dean of Computer Science at S I R T. Bhopal. He was Professor and Head Computer applications at M A N I T. Bhopal. He has 42 years of teaching and research experience. He has published more than 90 papers in National and International journals. He has written two books and guided 12 candidates for Ph D degree and 3 more are currently working. His current research interests are software engineering, cross language information retrieval, data mining, and knowledge management. EmailId: prof.mkumar@gmail.com